\documentclass[aps,twocolumn,showpacs,preprintnumbers,floatfix ]{revtex4-2}

\usepackage{graphicx}  
\usepackage{subfigure}
\usepackage{multirow}

\usepackage{fancyhdr}
\usepackage{longtable}
\usepackage[OT1]{fontenc}
\usepackage{dcolumn}   

\usepackage{bm}        
\usepackage{amsfonts}  
\usepackage{amsmath}   
\usepackage{amssymb}   
\usepackage{xcolor}
\usepackage{soul}
\usepackage{nicefrac}

\newcommand{\dd}{\mathrm{d}}

\newcommand{\pwisein}{\left\{ \begin{array}{ll}}
\newcommand{\pwiseout}{\end{array}\right.}

\begin{document}

\title{Heterogeneity can markedly increase final  outbreak size in the SIR model of epidemics}

\author{Alexander \surname{Leibenzon}, Michael \surname{Assaf}}

\affiliation {\it Racah Institute of Physics, Hebrew University of Jerusalem, Jerusalem 91904, Israel}

\begin{abstract}  

We study the SIR model of epidemics on positively correlated heterogeneous networks with population variability, and explore the dependence of the final outbreak size on the network heterogeneity strength and basic reproduction number $R_0$ -- the ratio between the infection and recovery rates per individual. We reveal a critical value  $R_0^c$, above which the maximal outbreak size is obtained at zero heterogeneity, but below which, the maximum is obtained at finite heterogeneity strength. This second-order phase transition, universal for all network distributions with finite standardized moments indicates that, network heterogeneity can greatly increase the final outbreak size. We also show that this effect can be enhanced by adding population heterogeneity, in the form of varying inter-individual susceptibility and infectiousness. Notably, our results provide key insight as to the predictability of the well-mixed SIR model for the final outbreak size, in realistic scenarios.
\end{abstract}


\maketitle


\textit{Introduction.} The SIR (susceptible-infected-recovered) model~\cite{KermackMcKemndrick1927, anderson1992infectious, Hethcote2000} has been a topic of great interest during the past decades~\cite{Chowell2016}, and
is one of the most conceptually basic, yet powerful models that describes the spread of an infectious disease. The model includes three population classes: susceptible (${\cal S}$), infected (${\cal I}$) and recovered (${\cal R}$). A contact 
between ${\cal S}$ and ${\cal I}$ individuals can give rise to the infection of ${\cal S}$. Conversely, an infected individual can recover and move to the ${\cal R}$ class. Remarkably, this simple model provides an adequate description to a wide variety of infectious diseases including COVID-19 pandemic~\cite{Yang2021}. 

Many works dealing with the SIR model assume a well-mixed topology; i.e., each individual interacts with all others (or has the same number of contacts)~\cite{anderson1992infectious, Hethcote2000, PastorSatorras2015, Saeedian2017, Bohner2019}. While this assumption is valid in some limits, in realistic scenarios one has to account for each individual's connectivity and deal instead with a \textit{population network}. In recent years, there have been several works dealing with the SIR model on heterogeneous random networks, where different individuals have varying connectivity~\cite{Miller2011, Volz2007, Newman2002, Kenah2007, Noel2009, Meyers2005}. In most of these works a mean-field approach is taken; i.e., the stochastic nature of the interactions and discreteness of individuals are neglected. 
Indeed, there have been other works that accounted for the demographic stochasticity in the SIR model, and studied the final outbreak size distribution~\cite{Ball1986, Ball1993,keeling2008modeling,House2013,Allen2017,Miller2019,House2013,Hindes2022}.
But even in the absence of demographic noise, while several authors have studied  epidemic spreading  on heterogeneous networks~\cite{Newman2002, Meyers2005, Volz2007, Miller2011}, to the best of our knowledge the direct influence of the network topology on the final outbreak size has not been studied. Importantly, this may be  key  for predicting the outcome of such a disease, as we show that the well-mixed (fully-connected) setting does not necessarily provide an upper bound for the final outbreak size.

Here we discover a novel second-order phase transition in the maximal outbreak size as a function of the network heterogeneity. Intuitively one would think that as the network heterogeneity increases, the final outbreak size should decrease, and thus, the outbreak size is maximized at zero heterogeneity. This is indeed the case for large values of the basic reproduction number $R_{0}={\beta}/{\gamma}$, describing the ratio between the infection rate $\beta$ and recovery rate $\gamma$ per individual. However, it turns out that there exists a critical value of $R_0$, which we denote by $R_0^c$, below which the maximal outbreak size is obtained at \textit{nonzero} heterogeneity. Furthermore, as $R_0$ is decreased below $R_0^c$, the magnitude of heterogeneity which maximizes the final outbreak size is increased. Interestingly, by introducing population heterogeneity in the form of varying susceptibility and/or infectiousness across individuals~\cite{Hethcote1978,Becker1989,Novozhilov2008, Gomes2022, LloydSmith2005,Neipel2020,tkachenko2021time}, this effect is enhanced, and the phase transition moves to increasingly larger values of $R_0^c$. In contrast, we find that the value of $R_0^c$ decreases as the degree-degree correlation between neighboring nodes increases.
Finally, we show that this phase transition is universal where $R_0^c$ is independent on the network topology, as long as the degree distribution has  finite standardized moments. 
Importantly, our results provide key insight  as to the limits of applicability of the simplified well-mixed SIR model on real-life heterogeneous networks with respect to the outbreak size. 

\textit{SIR model on networks.}
In the SIR model the sum of susceptibles ${\cal S}$, infected ${\cal I}$ and recovered ${\cal R}$ is conserved: ${\cal S}+{\cal I}+{\cal R}=N$. Here, $N$ represents to network size, i.e., the number of agents spreading the infection. Below,  we use  concentrations  of susceptibles, $S={\cal S}/N$, infected, $I={\cal I}/N$, and recovered, $R={\cal R}/N$. Denoting $R_{0}={\beta}/{\gamma}$, rescaling time $t\rightarrow \gamma t$, and assuming a well-mixed setting, in the limit of $N\gg 1$ the dynamics read:
\begin{equation}\label{MF}
\vspace{-1.0mm}\dot{S} = -R_{0} IS,\quad \dot{I} = R_{0} IS - I,\quad \dot{R}=I.
\vspace{-1.0mm}
\end{equation}
Notably, Eq.~\eqref{MF} ignores demographic noise, whose relative magnitude scales, in general, as $N^{-1/2}\!\ll\! 1$~\cite{assaf2010extinction,assaf2017wkb}. Moreover, in deriving Eq.~\eqref{MF} we used a fully-connected network, where each individual interacts with all others.

We now account for network heterogeneity by considering a population network, where each node represents an individual who can be either susceptible, infected or recovered, and edges between nodes represent interactions between them. We follow the formalism developed by Miller~\cite{Miller2011} and define $p(k)$ as the network degree distribution. Namely, $p(k)$ is the probability for a node to have $k$ neighbors. We furthermore assume that the network has positive degree-degree correlations~\cite{Moreno2003}, see below.

Let us denote $\theta(t)$ as the probability that a random edge has not transmitted an infectious contact up to a time $t$. This definition is equivalent to the probability that a node of degree $1$ is still susceptible at time $t$~\cite{Volz2007}. Thus, the probability of an individual node with $k$ neighbors to remain susceptible at time $t$ is given by $\theta^k$. As a result, the fraction of susceptibles at time $t$ is given by  
\vspace{-2mm}
\begin{equation}\label{S_miller}
S(t) = \sum_{k=0}^{\infty} p(k)\theta\left( t, \sigma \right)^k\equiv\psi(\theta,\sigma).\vspace{-2mm}
\end{equation}
Here, $\sigma$ is the standard deviation of the network degree distribution, $\sigma^2=\sum k^2 p(k)-k_0^2$, where $k_0=\sum_k k p(k)$, is the distribution's mean. Notably,  $\psi(\theta,\sigma)$ is the probability generating function of $p(k)$; its derivatives with respect to $\theta$ at $\theta=0$  provide the complete distribution, $p(k)$, while the derivatives at $\theta=1$ provide the distribution's moments; e.g., $k_0 = \partial_{\theta}\psi|_{\theta=1}$. While $\psi$ depends on the entire distribution $p(k)$, we have added an explicit dependence on $\sigma$, the heterogeneity strength, since we focus on the dependence of the final outbreak size on $\sigma$.

We now derive the governing equation for $\theta(t,\sigma)$ in order to obtain $S_{\infty}=\psi(t\to\infty)$, and the final outbreak fraction, $R_{\infty}=1-S_{\infty}$, where $S_{\infty}$ is the final susceptible fraction.
Below we set $\gamma=1$, such that time is measured in units of $\gamma^{-1}$, and rescale $\beta\to \beta/k_0$, such that $\beta$ now denotes the infection rate of a suscepetible node per infected neighbor. 
Defining an auxiliary variable $\phi(t)$ as the probability that a node $v$ is infectious but has not transmitted the disease to its neighbor $u$,  $\phi$ denotes the fraction of all $v\!-\!u$ edges in the network where $v$ is infected but has not (yet) directly infected $u$. Thus, $\dot{\theta}=-\beta\phi$~\cite{Miller2011}. 

The dynamics of $\phi(t)$ satisfies: $\dot{\phi}=-(\beta+1)\phi-\dot{h}$.
Here, $\phi$ decreases when the neighbor $u$ is infected from $v$ at rate $\beta\phi$, or when node $v$ is recovered at a rate of $\gamma\phi=\phi$. 
On the other hand, $\phi$ increases when a susceptible node $v$ becomes infected. Here, $h(t)$ is the probability that $v$ remains susceptible, and thus, 
$-\dot{h}(t)$ is the rate at which $v$ becomes infected from any of its neighbors except $u$. 
Accounting for positive degree-degree correlations, the probability that a neighbor of a degree-$k^{\prime}$ node has degree $k$, i.e., the two-point degree correlation function, satisfies: $p(k|k^{\prime})=(1-\alpha)kp(k)/k_0 + \alpha \delta_{k,k^{\prime}}$~\cite{Moreno2003}, where $\alpha$ measures the correlation strength~\footnote{For $\alpha=0$ we recover the result of random networks~\cite{Feld1991}.}. Therefore, $h(t)=\sum_{k^{\prime}=0}^{\infty}\sum_{k=0}^{\infty}p\left(k|k^{\prime}\right)p\left(k^{\prime}\right)\theta^{k-1}=(1-\alpha)k_0^{-1}\partial_\theta \psi\left(\theta, \sigma\right) + \alpha \theta^{-1} \psi\left(\theta, \sigma\right)$.
This derivation yields $\dot{\phi}=(1+1/\beta)\dot{\theta}-\dot{h}$, which can be integrated over time, using the fact that $\phi(0)=0$, $\theta(0)\simeq 1$ and $h(0)=1$. As a result,
\begin{equation}\label{theta_miller}
\dot{\theta}=1-\left(1+\frac{R_{0}}{k_{0}}\right)\theta+\frac{R_{0}}{k_{0}}\left(\frac{1-\alpha}{k_{0}}\partial_{\theta}\psi + \frac{\alpha}{\theta} \psi\right),
\end{equation}
where we have used the definition of $R_0=k_0\beta$.
This is a first-order nonlinear differential equation, which strongly depends on the network topology and degree correlations. While its time-dependent solution can be found numerically, we here study its steady-state solution, $\theta_{\infty}\equiv \theta(t\to\infty)$.
Indeed, putting $\dot{\theta}=0$ in Eq.~\eqref{theta_miller} we find:
\begin{equation}\vspace{-2mm}\label{theta_inf_varying_sigma}
\theta_{\infty}(\sigma) = \frac{R_{0}}{R_{0}+k_{0}}\left[\frac{1-\alpha}{k_{0}}\partial_{\theta}\psi+\frac{\alpha}{\theta_{\infty}}\psi + \frac{k_{0}}{R_{0}}\right]_{\theta=\theta_\infty}.
\end{equation}
Note that, for $\alpha=0$ the results of~\cite{Miller2011} are recovered. 

\begin{figure}[t]
\centering \includegraphics[scale=1.05]{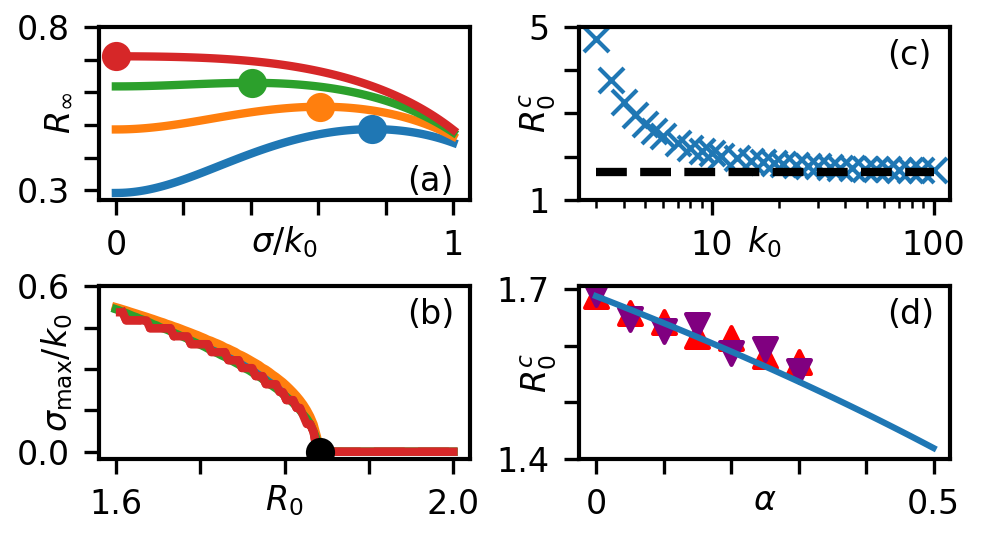}
    \vspace{-9mm}
	\caption{(a) A numerical solution of $R_\infty$ versus the network's COV, $\sigma / k_0$, for a bimodal network, see text: blue, orange, green, and red lines represent $R_0=1.3, 1.5, 1.7, 1.9$ respectively. (b) $\sigma_{\max} / k_0$, which maximizes  $R_\infty$ versus $R_0$: blue, orange, green, and red lines respectively show  bimodal, symmetric beta, gamma and uniform distributions. Here $R_0^c\simeq 1.84$. 
    (c) $R_0^c$ versus $k_0$; dashed line is the asymptotic value of $(3/2)\ln 3$. (d) $R_0^c$ versus $\alpha$; theoretical solution of~\eqref{max_R_inf-relation} (solid line) is compared with simulations of bimodal (upper triangles) and gamma (lower triangles) distributions. In all panels $N=10^4$ and in (a-b) $k_0=20$, while in (d) $k_0=100$.}
\label{fig:fig1}
\vspace{-3mm}
\end{figure}

\textit{Maximal outbreak size.}
Equation~(\ref{theta_inf_varying_sigma}) can be numerically solved  for various network topologies, $p(k)$, having mean $k_0$ and standard deviation $\sigma$. 
An example for the dependence of $R_{\infty}$ on $\sigma$, for various values of $R_0$, can be seen in Fig.~\ref{fig:fig1}(a) where we have used a bimodal network, with $p(k)={1}/{2}\left(\delta_{k,k_0-\sigma}+\delta_{k,k_0+\sigma}\right)$.
Remarkably, as $R_0$ is lowered below some threshold $R_0^c$, the maximum of $R_{\infty}$ shifts from $\sigma=0$ to $\sigma>0$. That is, while for $R_0>R_0^c$, the final outbreak size is maximized when the network is homogeneous, for $R_0<R_0^c$ the maximum is obtained at finite heterogeneity. This result is counter intuitive. As $\sigma$ is increased, the final outbreak size should decrease, as nodes with very high degree become more abundant. Due to their high degree, these nodes get infected (and recovered) much quicker than lower-degree nodes, which causes a more rapid decrease in the effective infection rate per individual, and correspondingly, in the final outbreak size, compared to the homogeneous case. Yet,$\!$ here we show that this phenomenon is not universal, but rather depends on the underlying value of $R_0$.  

We have studied the dependence of the threshold, $R_0^c$, on the network's degree distribution. In Fig.~\ref{fig:fig1}(b), we plot the value of the coefficient of variation (COV), $\sigma/k_0$,  which maximizes the final outbreak size, for bimodal, symmetric beta, gamma and uniform distributions, versus $R_0$, for $k_0=20$ and $\alpha=0$.
The fact that all curves collapse indicates that  $R_0^c$ is universal and is independent on the particular details of the network details, see below.

To find $R_0^c$  we realize that at the threshold, $R_0=R_0^c$, the maximum of $R_{\infty}$ is obtained exactly at $\sigma=0$, namely ${\dd R_{\infty}}/{\dd\sigma}|_{\sigma=0}=0$. Above $R_0^c$ this derivative is negative, whereas below $R_0^c$ the maximum is obtained for $\sigma>0$, see Fig.~\ref{fig:fig1}(a).  
Differentiating $R_{\infty}=1-\psi(\theta_\infty,\sigma)$ with respect to $\sigma$, using Eqs.~(\ref{S_miller}) and (\ref{theta_inf_varying_sigma}), and demanding that the derivative ${\dd R_{\infty}}/{\dd\sigma}$ be zero at $\sigma=0$, we arrive at
\begin{equation}\label{max_R_inf-relation}
\hspace{-2.5mm} k_0\!+\!\frac{k_{0}^2}{R_{0}^{c}}\!=\! \left.\!\left[\!(1\!-\!\alpha)\!\left(\!\partial_{\theta\theta}\psi\!-\!\frac{\partial_{\theta}\psi\partial_{\sigma\theta}\psi}{\partial_{\sigma}\psi}\!\right)\!-\!\frac{\alpha k_{0}}{\theta^2}\psi\!\right]\!\right|_{\theta=\theta_{\!\infty},\sigma=0}\!\!\!.
\end{equation}
This is an exact algebraic equation, whose solution provides $R_0^c$. In general it can be solved numerically, whereas analytical progress can be made for $k_0\gg 1$. Here we seek for the solution perturbatively by assuming $\theta_\infty = 1 - \epsilon$ with $\epsilon={\cal O}(k_0^{-1})\ll 1$ (to be verified a-posteriori).

First, we establish a connection between $\epsilon$ and $R_0^c$ by plugging $\theta_\infty\! =\! 1 - \epsilon\,$ into~\eqref{theta_inf_varying_sigma}, and putting $\sigma\!=\!0$, i.e., using a homogeneous distribution, $p(k)=\delta_{k,k_0}$. Keeping leading order terms we arrive at  $ \epsilon k_{0} \simeq R_{0}^c\left[1-\exp({-\epsilon k_{0}})\right]$, the solution of which is given via the Lambert W-function
\begin{equation}\label{epsilon-for-theta_inf} 
\epsilon=k_0^{-1}\{R_{0}^c+W_{0}[-R_{0}^c\exp(-R_{0}^c)]\}.
\end{equation}
Going back to Eq.~\eqref{max_R_inf-relation}, for $k_0\gg 1$, $\psi(\theta_\infty)=\theta_\infty^k$ can be approximated as $\theta_\infty^k=\exp(-\epsilon k)$, with ${\cal O}(k \epsilon^2)\ll 1$ corrections in the exponent. Thus, the two terms $\partial_{\theta}\psi(\theta,\sigma)$ and $\partial_{\theta\theta}\psi(\theta,\sigma)$ evaluated at $\theta=\theta_\infty$ and
$\sigma=0$, read:
\begin{equation} 
        \partial_{\theta\theta}\psi \!=\! e^{-\epsilon k_{0}}k_{0}^{2}\left[1 \!+\! \mathcal{O}(\epsilon)\right],\;\;
        \partial_{\theta}\psi \!=\! e^{-\epsilon k_{0}} k_0\left[1 \!+\!\mathcal{O}(\epsilon) \right].
        \label{sum_Rc}
\end{equation}

Notably, the terms involving derivatives with respect to $\sigma$ in Eq.~\eqref{max_R_inf-relation} are more involved as one has to use the definition of $\psi$ from Eq.~\eqref{S_miller}. To proceed, we write
\begin{equation} \label{inter}
        \frac{\partial_{\sigma\theta}\psi\left(\theta,\sigma\right)}{\partial_{\sigma}\psi\left(\theta,\sigma\right)}
        =k_{0}-\partial_{\epsilon}\ln\partial_{\sigma}\left\langle e^{-\epsilon\left(k-k_{0}\right)}\right\rangle+\mathcal{O}\left(1\right),
\end{equation}
where this expression has to be evaluated at $\theta=\theta_{\infty}$ and $\sigma=0$. Here, we added  $k_0$, and subtracted $k_0$ by subtracting $k_0$ from $k$ in the exponent. The term in the brackets is (up to a minus sign) the generating function of the central moments (around the mean) $\{\mu_n\}_{n=0}^{\infty}$. Taylor-expanding in powers of $\epsilon(k-k_0)$, we find:
$\left\langle e^{-\epsilon\left(k-k_{0}\right)}\right\rangle=\sum_k p(k)-\epsilon\sum_k p(k)(k-k_0)+(\epsilon^2/2)\sum_k p(k)(k-k_0)^2-(\epsilon^3/6)\sum_k p(k)(k-k_0)^3+...=1+(\epsilon^2/2)\sigma^2-(\epsilon^3/6)\mu_3+...$, where $\mu_1=0$ and $\mu_2=\sigma^2$. For  $p(k)$ with finite standardized moments, $\tilde{\mu}_n$, one can show that $\mu_n=\sigma^n\tilde{\mu}_n$. As a result, plugging this series back into Eq.~\eqref{inter}, all terms with powers of $\sigma$ greater than $2$ vanish, since we set $\sigma=0$ after the differentiation, and one finally obtains:
$\partial_{\sigma\theta}\psi\left(\theta,\sigma\right)/\partial_{\sigma}\psi\left(\theta,\sigma\right) = k_{0}-2/\epsilon+\mathcal{O}\left(1\right)$.
Plugging this along with Eq.~\eqref{sum_Rc}  into~\eqref{max_R_inf-relation}, and using Eq.~\eqref{epsilon-for-theta_inf}, in the leading order of $k_0\gg1$ the critical $R_{0}^c$ is found to be
\begin{equation}\label{R0c}
    R_{0}^c=[(3-2\alpha)/(2-2\alpha)]\ln(3-2\alpha).
\end{equation}
For uncorrelated networks, $\alpha=0$, we find $R_{0}^c=(3/2)\ln3\simeq 1.648$. Plugging $R_{0}^c$ into Eq.~\eqref{epsilon-for-theta_inf} verifies a-posteriori that $\epsilon={\cal O}(k_0^{-1})$. We have checked that as $k_0$ is increased, the numerical value of $R_0^c$ approaches our theoretical prediction given by Eq.~\eqref{R0c}, see Fig.~\ref{fig:fig1}(c)~\footnote{For $k_0=\mathcal{O}(1)$ our derivation is invalid since $\epsilon={\cal O}(1)$, and in addition, stochastic effects become dominant, such that $R_0^c$ rapidly grows as $k_0$ is decreased;  see Fig.~\ref{fig:fig1}(c).}.

To verify our results we ran Gillespie simulations~\cite{Gillespie1977} on correlated, bimodal and gamma distributed networks, of size $N=10^4$ and mean degree $k_0=100$. To achieve a given correlation $\alpha$, for each degree-$k$ node having initial $k$ stems, a fraction $\alpha$ of its stems were connected to stems of other degree-$k$ nodes, while the rest were connected randomly, as in the configuration model~\cite{Molloy1995}. This algorithm creates a network with correlation  $\alpha$ for small $\alpha$, while it tends to lose accuracy as $\alpha$ grows, due to finite size effects.
In Fig.~\ref{fig:fig1}(d) our theoretical prediction~(\ref{R0c}) is shown to agree well with simulations  at low $\alpha$'s. While we focus  on $\alpha > 0$ indicative of social networks~\cite{Newman2002-assortativity}, we checked that for $\alpha < 0$, $R_0^c$ grows as expected.

\begin{figure}[t]
    \includegraphics[scale=1.0]{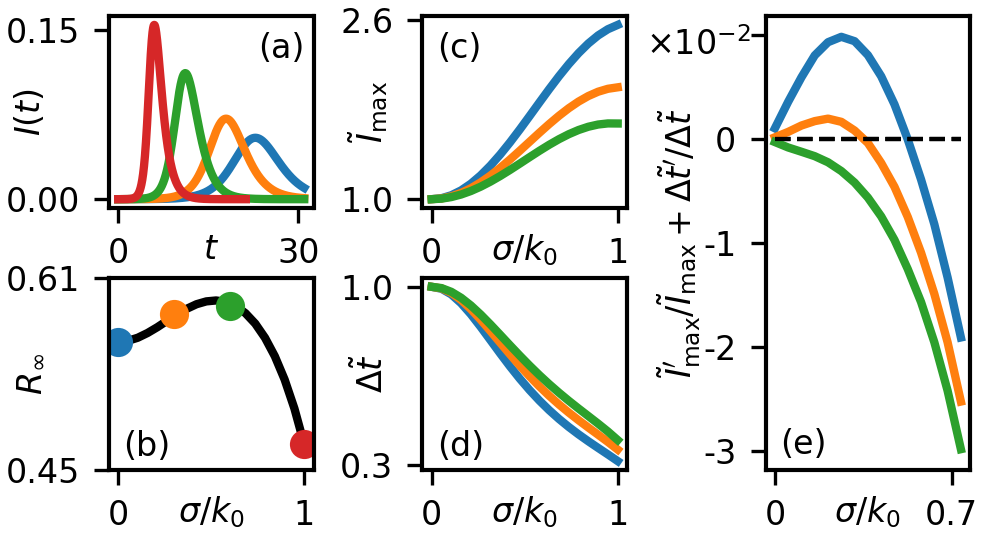}
   \vspace{-4mm}\caption{(a) Infected fraction $I(t)$ versus time, for $R_{0}=1.6$. Blue, orange, green and red lines represent COVs, ${\sigma}/{k_0}=0,0.3,0.6,1$, respectively. (b) Line shows the approximation for the final outbreak fraction $c I_{\max} \Delta t$ versus the COV, for the same network as in (a). Dots represent numerical integration over $I(t)$ from (a). 
   Panels (c), (d) and (e) respectively show $\tilde{I}_{\max}$, $\Delta \tilde{t}$ and $\tilde{R}_\infty'/\tilde{R}_\infty=I_{\max}^{\prime}/I_{\max}+\Delta t^{\prime}/\Delta t$ versus the COV,  where  prime denotes differentiation with respect to $\sigma$.
   In all panels we use a bimodal distribution with $k_0=20$ and $\alpha=0$, and in (c)-(e) blue, orange and green lines represent $R_0=1.65,1.8,1.95$ respectively (here $R_0=1.95$ is above $R_0^c$). \vspace{-7mm}
    \label{fig:fig2}}
    \end{figure}

What is the reason for the second-order phase transition  observed in Fig.~\ref{fig:fig1}(b)?
The total outbreak size satisfies  $R_\infty=\int_{0}^{\infty}I(t)\dd t$. Several examples of epidemic waves for various COV values are shown in Fig.~\ref{fig:fig2}(a). We propose to approximate $R_\infty$ as $R_{\infty}\simeq cI_{\max}\Delta t$,
where $I_{\max}$ is the maximal value of $I$ (that defines herd immunity), and $\Delta t$ is the typical wave's duration: the time interval during which $I$ is greater than a fraction $f$ (yet to be found) of $I_{\max}$, while $c$ is a constant.  For the distributions we have studied, $f$ and $c$ were found to satisfy $f\approx0.27$ and $c\approx0.785$ for a wide range of $R_0=\mathcal{O}(1)$ and $\sigma$ values. In Fig.~\ref{fig:fig2}(b) the approximate and exact solutions for $R_\infty$ agree well, for a bimodal networks~\footnote{The maximal relative error in Fig.~\ref{fig:fig2}(b) between the numerical values of $R_\infty$ and those obtained by the approximated formula with the fitted parameters was $<0.2\%$.}.  

To explain the appearance of a phase transition at $R_{0}^c$, we denote by $\tilde{I}_{\max}$ (and similarly for $\Delta \tilde{t}$) the ratio of $I_{\max}$ at given $\sigma$ and its value at $\sigma=0$, see Fig.~\ref{fig:fig2}(c)-(d), such that  $\tilde{R}_{\infty}=\tilde{I}_{\max}\Delta\tilde{t}$. Thus, we have $\tilde{R}_\infty'(\sigma)/\tilde{R}_\infty(\sigma)=\tilde{I}_{\max}'(\sigma)/\tilde{I}_{\max}(\sigma)+\Delta\tilde{t}'(\sigma)/\Delta\tilde{t}(\sigma)$. At $R_0>R_0^c$ we see from Fig.~\ref{fig:fig2}(e) that $\tilde{R}_\infty'(\sigma)/\tilde{R}_\infty(\sigma)$ is negative for any $\sigma$. Yet, as $R_c$ goes below $R_0^c$ a non-monotone regime appears, which gives rise to a maximum in $R_{\infty}$ at $\sigma>0$. 

This can be understood as follows. As the network heterogeneity strength $\sigma$ is increased, there are more very high degree nodes (hubs), which get infected first due to their high degree, and infect the entire network rapidly. This rapid epidemic spread causes $I$ to surge, but also causes the epidemic's duration $\Delta t$ to decrease. For low infection rates,  $R_0<R_0^c$,  increasing $\sigma$ initially causes the increase of $R_{\infty}$ as the increase of $I_{\max}$ cannot be balanced by the decrease of $\Delta t$, see Fig.~\ref{fig:fig2}(c)-(e). Notably, as $\sigma$ exceeds $\sigma_{\max}$ the rate of spread of the hubs is so rapid such that low-degree nodes are hardly infected, and thus, $R_\infty$ starts to decrease. Exactly at the onset of decrease of $R_{\infty}$, i.e. at $\sigma=\sigma_{\max}$, the disease spread rate is optimal such that the total number of infected nodes, $R_\infty$ is maximized.
Importantly, increasing $R_0$ has a similar effect to increasing $\sigma$. That is, when $R_0$ grows, the increase of $\sigma$ is no longer needed to increase the rate of disease spread. Thus, if $\sigma$ is also increased, one exceeds the optimal disease spread rate which yields a decline in $R_{\infty}$. Therefore, if at  $R_0<R_0^c$ the maximum of $R_{\infty}$ is obtained at $\sigma=\sigma_{\max}>0$, as $R_0$ is increased, $\sigma_{\max}$ shifts towards zero, as increasing $R_0$ is complementary to increasing $\sigma$.

\textit{Population heterogeneity.}
We now add variability across the population (population heterogeneity) and study its effect on the phase transition, by using the formalism of~\cite{Neipel2020} and modulating the infection rate $\beta$ by the mean population's susceptibility $\Bar{x}$, such that   $R_0\rightarrow\Bar{x}R_0$. While for  homogeneous populations $\Bar{x}(t) = 1$, for heterogeneous populations, $\Bar{x}$ decays in time, as the highly susceptible individuals get infected and recover relatively quickly thereby decreasing $\Bar{x}$. In the well-mixed case,  denoting by $s(x,t)$ the fraction of susceptibles having infection rate between $x$ to $x+d x$, the total fraction of susceptibles is $S(t) = \int^{\infty}_{0} s(x,t) \dd x$. Thus,  $s(x,t)$ satisfies $\partial_{t}s = - R_{0} x s I$, and the mean susceptibility becomes~\cite{Neipel2020}
\vspace{-4mm}
\begin{equation}\label{average-x-t} \Bar{x}(t) = S(t)^{-1} \int^{\infty}_{0} x s(x,t) \dd x.
\vspace{-2mm}\end{equation}
To find $\bar{x}(t)$, a new  time scale  $\tau\!=\!R_0R$ is defined, measuring the epidemic spreading. Thus, $\dd \tau / \dd t=R_0I$, such that $\partial_{\tau}s = - x s$, which yields: $s(x, \tau) = s_0(x) \exp(-\tau x)$. 
We incorporate  population heterogeneity by taking a gamma-distributed initial susceptibility, $s_0(x) \sim x^{-1+a}e^{-a x}$, with average 1 and standard deviation  $\sigma_p=a^{-1/2}$~\footnote{Naturally, other distributions of population heterogeneity  are also possible. Yet, the effect we describe is generic and is independent on the specific choice of distribution.}. With this distribution, $\Bar{x}$ given by Eq.~\eqref{average-x-t} decays in time as $\Bar{x} = \left(1+\tau\sigma_p^2\right)^{-1}$~\cite{Neipel2020}.

\begin{figure}[t!]
\centering
\includegraphics[scale=1]{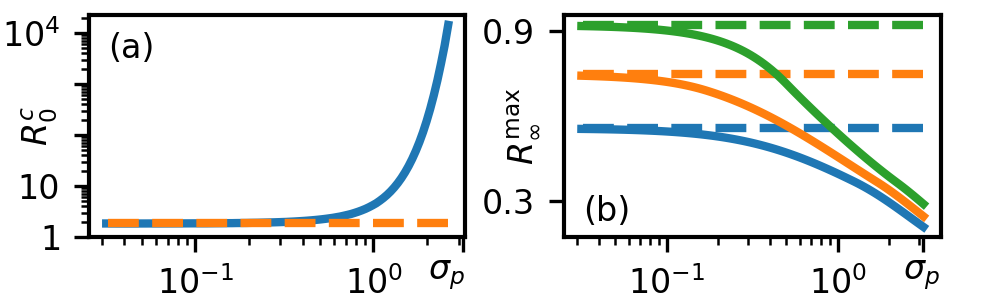}
    \vspace{-3mm}
	\caption{(a) Critical basic reproduction number $R_{0}^c$ versus population heterogeneity strength $\sigma_p$. Dashed line is $R_{0}^c\approx1.84$, the asymptotic value of $R_0^c$ at $\sigma_p\to 0$. (b)  Maximal outbreak size, $R_{\infty}^{\max}$ versus $\sigma_p$, for $R_0=1.5,2,3$  (blue, orange and green lines, respectively). Dashed lines are the asymptotic values for $\sigma_p\rightarrow 0$. In both panels $k_0=20$ and $N=10^4$.}
 \vspace{-4mm}
\label{fig:fig3}
\end{figure}

To combine network and population heterogeneity,
we introduce the dynamical infection rate $\beta(t)=\Bar{x}\beta=\Bar{x}R_0/k_0$ with $\Bar{x}$ given by Eq.~\eqref{average-x-t}. For heterogeneous networks, $\theta$ and $\phi$ are connected via: $\dot{\theta} = -\beta\phi$. Using the equation for $\phi$ defined above Eq.~\eqref{theta_miller}, putting $\beta \to \bar{x}(t)\beta$, and differentiating $\dot{\theta}$ with respect to time, we arrive at
\begin{equation}\label{theta_dot_dot_combined}
\ddot{\theta} \!=\! \dot{\theta} \! \left\{\! \frac{\dot{\Bar{x}}}{\Bar{x}} \!+\! \Bar{x}\frac{R_0}{k_0} \!\left[\! (\!1\!-\!\alpha\!)\frac{\partial_{\theta\theta}\psi}{k_0} \!-\! \alpha \!\left(\! \frac{\psi}{\theta^2} \!-\! \frac{\partial_{\theta}\psi}{\theta} \!\right) \!-\! 1 \!\right]\!\! - \!1 \!\right\}\!\!,
\end{equation}
where we have assumed a correlation strength $\alpha$. The validity of Eq.~\eqref{theta_dot_dot_combined} can be checked in two limits. In the limit of homogeneous population, $\Bar{x}\to 1$ and Eq.~\eqref{theta_miller} is restored upon integration over time. In the well-mixed limit, $k_0\simeq N\gg 1$,  $\theta=1-\mathcal{O}(k_0^{-1})$; here a proportion of $\mathcal{O}({I}/{k_0})$ of edges emanating from each node transmits the infection from a still infected node~\cite{Miller2011}. Thus, $\phi=1-\mathcal{O}({I}/{k_0})$, $\psi^\prime(\theta)={k_0 S}/{\theta}$, and $\dot{S}=-\psi^\prime(\theta)\dot{\theta}=-R_0 \Bar{x} S I +\mathcal{O}({S I}/{k_0})$, which coincides in the leading order with the well-mixed SIR model under population heterogeneity~\cite{Neipel2020,tkachenko2021time}.

To find $R_0^c$ under both population and network heterogeneity, we numerically compute the steady-state solution of Eq.~\eqref{theta_dot_dot_combined}~\footnote{Here, unlike the homogeneous case~\eqref{theta_inf_varying_sigma}, integration cannot be performed, as $\bar{x}$ explicitly depends on time.}, which allows finding $R_{\infty}(R_0)$. Here, as $\Bar{x}$ decreases over time, the  effective disease spread rate, $\Bar{x}R_0$, decreases, which can be compensated by more highly connected nodes. Thus, $R_{0}^c$ increases as population heterogeneity increases, namely as  $\sigma_p$ increases. This is demonstrated  for a bimodal network in Fig.~\ref{fig:fig3}(a).


\textit{Discussion.}
We have discovered a previously unknown phase transition in the maximum value of the final outbreak size $R_{\infty}^{\max}$, as function of the network heterogeneity strength, $\sigma$, as $R_0$ crosses a threshold of $R_0^c$.  While for $R_0>R_0^c$, $R_{\infty}^{\max}$ is obtained at $\sigma=0$, for $R_0<R_0^c$, $R_{\infty}^{\max}$ is obtained at $\sigma>0$. This counter-intuitive result stems from an intricate balance between the increase in the peak and decrease in the duration of the epidemic wave, as the network heterogeneity grows. We  also showed that population heterogeneity and degree correlations between  neighboring nodes strongly affect the value of $R_0^c$. 


What are the implications of this phase transition for realistic scenarios? For diseases such as the smallpox, monkeypox, diphtheria or COVID-19, $R_0>2$ is above $R_0^c$~\cite{Gani2001,Grant2020,Truelove2020,Billah2020,Liu2022}. Here, the prediction of the well-mixed SIR model gives an upper bound for $R_{\infty}$. Yet, for $R_0<R_0^c$, taking the well-mixed SIR prediction as an upper bound may be erroneous; e.g., for seasonal influenza ($R_0=1.28$~\cite{Biggerstaff2014}), $\sigma_{\max} / k_0 \simeq 0.857$ for a gamma-distributed network with  $k_0=20$. This yields $R_{\infty}\simeq 0.466$,  higher by  $\sim$16\% than the well-mixed prediction, $R_{\infty}\simeq 0.403$. Notably, for positively correlated networks, $R_0^c$ decreases, whereas adding population heterogeneity decreases $R_0^c$. Yet, in Fig.~\ref{fig:fig3}(b) the decrease in $R_{\infty}^{\max}$ for all values of $\sigma$, due to population heterogeneity, supersedes the increase in $R_{\infty}^{\max}$ due to network heterogeneity. Thus, while evaluating $R_0^c$ and $R_{\infty}^{\max}$ in realistic scenarios is highly non-trivial, it may provide important insight as to the outcome of the epidemics in the worst-case scenario.



\textit{Acknowledgements.} AL and MA acknowledge support from the ISF
grant 531/20.

\bibliography{bib_file}

\end{document}